\documentstyle[12pt]{article}

\date{June 1995\\LTP-043-UPR}
\title{Remarks on the Coulomb and Covariant Gauges in Finite Temperature QED}
\author{J. C. D'Olivo\thanks{
Partially supported by
Grant No. DGAPA-IN100694}\\
	Instituto de Ciencias Nucleares\\
	Universidad Nacional Aut\'{o}noma de M\'{e}xico\\
	Apartado Postal 70-543, 04510 M\'{e}xico, D.F., M\'{e}xico
	\and Jos\'{e} F. Nieves\thanks{
Partially supported by the
US National Science Foundation Grant PHY-9320692}\\
	Laboratory of Theoretical Physics\\
	Department of Physics, P. O. Box 23343\\
	University of Puerto Rico\\
R\'{\i}o Piedras, Puerto Rico 00931-3343
}

\begin{document}

\maketitle

\begin{abstract}

We compare the use of the Coulomb gauge in finite
temperature QED
with a recently proposed prescription for covariant gauges,
in which only the transverse photon degrees
of freedom are thermalized.   Using the Landau rule as a guide,
we clarify the relation between the
retarded electron self-energy and the elements of the
self-energy matrix in the real-time formulation of .
The general results are illustrated  by means of the
one-loop expressions for the electron self-energy in a QED plasma.

\end{abstract}

In a theory with fermions and scalars only,  the
real-time formulation of  Finite Temperature Field Theory (FTFT)
\cite{holandeses, canonical} is quite straightforward.
However,  the situation becomes more involved in gauge theories
like QED. For covariant  gauges, the traditional  approach
has been to assume that
all the degrees of freedom of the gauge bosons are
in thermal equilibrium\cite{bernard}.
Recently, Landshoff and Rebhan (LR)\cite{landshoff}
showed that it is possible and even
simpler to assume that only the physical transverse
components of the gauge field are thermalized.
The price is that the operator averages
cannot be expressed as traces and therefore
some  formulas of the standard formalism do not apply anymore.
Here,  we elaborate this point
by considering the relation between the physical
(retarded) fermion self-energy and the elements
of the self-energy  matrix calculated with the Feynman rules of the theory.
The formulas  we write  for the dispersive and absorptive parts of the
effective
self-energy satisfy the Landau rule and  are the
appropriate ones for those situations
where statistical averages cannot be represented by a trace.

Within the real-time formalism,
the self-energy
of a fermion in a thermal background
is a $2\times 2$ matrix, whose elements are defined by
\begin{eqnarray}\label{defsigma}
i\Sigma_{21}(z - y)_{\alpha\beta} & = & -\langle\eta_\alpha(z)
\overline\eta_\beta(y)\rangle\,,\nonumber\\
i\Sigma_{12}(z - y)_{\alpha\beta} & = & \langle\overline\eta_\beta(y)
\eta_\alpha(z)\rangle\,,\nonumber\\
-\Sigma_{11}(z - y) & = & \Sigma_{21}(z - y)\theta(z^0 - y^0)
+ \Sigma_{12}(z - y)\theta(y^0 - z^0)\,,\nonumber\\
-\Sigma_{22}(z - y) & = & \Sigma_{21}(z - y)\theta(y^0 - z^0)
+ \Sigma_{12}(z - y)\theta(z^0 - y^0)\,,
\end{eqnarray}
where $\eta$ and $\overline\eta$ are the fermion source
fields.  In terms of them the interaction Lagrangian
is
\begin{equation}\label{Lint}
L_{\mbox{int}} = \overline\psi\eta + \overline\eta\psi\,,
\end{equation}
and in particular,  for QED  $\eta = -eA\mkern -9.0mu\psi$.  The angle
brackets in Eq.~(\ref{defsigma}) stand for the statistical average
which, for any operator ${\cal O}$ is defined by
\begin{equation}\label{defaverage}
\langle{\cal O}\rangle = \frac{\sum_i\langle i|\rho{\cal O}|i\rangle}
{ \sum_i\langle i|\rho|i\rangle}\,,
\end{equation}
where
\begin{equation}\label{rho}
\rho = e^{-\beta H + \sum_{A}\alpha_A Q_A}\,.
\end{equation}
$H$ is the Hamiltonian
of the system, the quantities $Q_A$ are the (conserved) charges
that commute with $H$,  $1/\beta$ is the temperature $T$,
and the $\alpha_A$ are the chemical
potentials that characterize the composition of the background.

In a theory without gauge fields
the sums in Eq. (\ref{defaverage})
are over all the states of the system and are unambiguous.
The same is true for QED in the Coulomb gauge.
In this case,  the Hilbert space contains only physical states
and  the unphysical photon degrees of freedom  disappear,
along with the associated question of wether they have a thermal
distribution or not.  Then,  the photon propagator takes the form
\begin{equation}\label{photonpropcoul}
\Delta_{ab}^{\mu\nu}(k) = (-S^{\mu\nu})[\Delta^{(0)}_{ab}(k)
+ \Delta^{(T)}_{ab}(k)]\,,
\end{equation}
with $S_{\mu\nu}$ given by
\begin{equation}\label{Smunu}
S_{\mu\nu} = g_{\mu\nu} + \frac{1}{\kappa^2}k_\mu k_\nu -
\frac{\omega}{\kappa^2}(u_\mu k_\nu + k_\mu u_\nu)\,,
\end{equation}
where $\omega = k\cdot u$ and $\kappa = \sqrt{\omega^2 - k^2}$
are the energy and the magnitude of the 3-momentum
$\vec{\kappa}$ of the photon in the frame where the medium is at rest.
We have introduced the vector $u^\mu$ representing the velocity
4-vector of the background, with components
 $(1,\vec{0})$ in its own rest frame.
In Eq.~(\ref{photonpropcoul}),
\begin{equation}\label{auxpropagator1}
\Delta^{(0)}_{ab}(k) = \left(
\begin{array}{cc}
\frac{1}{k^2 + i\epsilon} & -2\pi i\delta(k^2)\theta(-k\cdot u) \\
& \\
-2\pi i\delta(k^2)\theta(k\cdot u) & \frac{-1}{k^2 - i\epsilon}
\end{array}\right)\,,
\end{equation}
and
\begin{equation}\label{auxpropagator2}
\Delta^{(T)}_{ab}(k) = -2\pi i\delta(k^2)\frac{1}{e^{\beta\left|k\cdot
u\right|} - 1}\left(
\begin{array}{cc}
1 & 1 \\
1 & 1
\end{array}\right)\,.
\end{equation}
It is useful to observe that
\begin{equation}\label{polarizationsum}
\left.S_{\mu\nu}\right|_{\omega = \kappa} =
-\left.\sum_{\lambda = 1,2}{\epsilon_\mu(k,\lambda)
\epsilon_\nu(k,\lambda)}\right|_{\omega = \kappa}\,,
\end{equation}
where the polarization vectors are given by
$\epsilon^\mu(k,\lambda) = (0,\vec{e}(k,\lambda))$,
with $\vec{e}(k,\lambda)\cdot\vec{k} = 0$.

The situation is not so obvious in a covariant gauge,
because the set of physical states does not span the whole space.
In the traditional approach the sum is made over
a complete set of states and the unphysical
degrees of freedom  acquire a thermal part.
The covariant  photon propagator has  the same form
as in Eq.~(\ref{photonpropcoul}), but with  $S_{\mu\nu}$
replaced by  the tensor $C_{\mu\nu}$ whose  explicit
expresion  depends on the gauge in which the
theory is quantized. For example, in the Feynman gauge
 $C_{\mu\nu} = g_{\mu\nu}$.
On the other hand, according to the prescription
of LR, the sums in Eq (\ref{defaverage})
involve only  physical states,
even when the theory is formulated in a covariant gauge.
We will refer to this approach as the {\em mixed gauge},
and in it the  thermal part of the  photon propagator is the
same as in the Coulomb gauge,
while the zero-temperature term has the structure
that corresponds to a covariant gauge:

\begin{equation}\label{photonpropmixed}
\Delta_{ab}^{\mu\nu}(k)^{(mix)} = (-C^{\mu\nu})\Delta^{(0)}_{ab}(k)
+ (-S^{\mu\nu})\Delta^{(T)}_{ab}(k)\,.
\end{equation}
In that manner,  this approach
attempts  to combine the simplicity
of covariant gauges for the vacuum
part with the advantages of the (noncovariant)
Coulomb gauge for the
temperature-dependent terms and in principle, may be more
convenient for calculational purposes.
Nevertheless,  as we discuss next, the fact that
thermal averages are made by summing
over a subset of states of the Hilbert space,  has consequences
that cannot be ignored.

In a medium, the effective field
equation for a fermion with momentum
$p^\mu = (\varepsilon,\vec P)$ is \footnote{ Eq.~(\ref{fermionfiledeq})
can be derived  from the functional
derivative
of the effective action, by a procedure similar
to the one described in Ref.~\cite{canonical} for a scalar particle
\cite{footnote1}.}
\begin{equation}\label{fermionfiledeq}
(\rlap / p - m - \Sigma_{eff})\psi = 0\,,
\end{equation}
where
\begin{equation}\label{sigmaeffgeneral}
\Sigma_{eff}(p) = \Sigma_{11}(p) + \Sigma_{12}(p)\,.
\end{equation}
As seen from Eq.~(\ref{defsigma}) $\Sigma_{eff}$
corresponds to  the retarded self-energy.
Denoting by $\mbox{Re}\,\Sigma_{11}$ and
$\mbox{Im}\,\Sigma_{11}$
the dispersive and absorptive parts of
$\Sigma_{11}$:
\begin{eqnarray}\label{eq2.18}
\mbox{Re}\,\Sigma_{11} & = & \frac{1}{2}(\Sigma_{11} +
\gamma_0\Sigma_{11}^\dagger\gamma_0)\,,\nonumber\\
\mbox{Im}\,\Sigma_{11} & = & \frac{1}{2i}(\Sigma_{11} -
\gamma_0\Sigma_{11}^\dagger\gamma_0)\,,
\end{eqnarray}
with a similar decomposition for $\Sigma_{eff}$,
Eq.~(\ref{sigmaeffgeneral}) is equivalent to
\begin{eqnarray}\label{sigmaeffgeneral2}
\Sigma_r (p)& = & \mbox{Re}\,\Sigma_{11}(p)\,,\nonumber\\
\Sigma_i (p) & = & \mbox{Im}\,\Sigma_{11}(p)  - i \Sigma_{12}(p)\,.
\end{eqnarray}
We have used the fact that $\Sigma_{12}$ is purely
absorptive, as follows from its definition in (\ref{defsigma}).
Now we verify that  these formulas are related correctly
by the spectral representation
as required on the basis of
fundamental principles\cite{fetter}.
By using the integral representation of the step function,
the following expressions are easily derived from Eq.~(\ref{defsigma}):
\begin{eqnarray}\label{disprel1}
\mbox{Re}\,\Sigma_{11}(p) & = &
\frac{1}{2\pi i}{\cal P}
\int d\varepsilon^\prime\frac{\Sigma_{21}(\varepsilon^\prime,\vec P) -
\Sigma_{12}(\varepsilon^\prime,\vec P)}
{\varepsilon - \varepsilon^\prime}\,,\nonumber\\
\mbox{Im}\,\Sigma_{11}(p) & = &
\frac{i}{2}[\Sigma_{21}(p) + \Sigma_{12}(p)]\,,
\end{eqnarray}
where  $\varepsilon = p\cdot u$. The second of these equations
implies that $\Sigma_i$,
determined according to  Eq.~(\ref{sigmaeffgeneral2}),
can be also computed by  means of
\begin{equation}\label{sigmaeffabs}
\Sigma_i(p) = \frac{i}{2}[\Sigma_{21}(p) - \Sigma_{12}(p)]\,,
\end{equation}
that  substituted in the formula for $\mbox{Re}\,\Sigma_{11}(p)$,  gives
\begin{equation}\label{disprel3}
\Sigma_r(p) = \frac{-1}{\pi}{\cal P}\int
d\varepsilon^\prime\frac{\Sigma_i(\varepsilon^\prime,\vec P)}
{\varepsilon - \varepsilon^\prime}\,,
\end{equation}
or, equivalently
\begin{equation}\label{landaurule}
\Sigma_i(p) = \mbox{Im}\,\Sigma_r(\varepsilon + i\epsilon,\vec P)\,.
\end{equation}
Eqs.(\ref{disprel3}) and (\ref{landaurule}) are just the
statement of the Landau rule within the present context,
and as the above reasoning shows, they will always
be satisfied if the retarded self-energy is calculated from
Eqs.~(\ref{sigmaeffgeneral2}) or (\ref{sigmaeffabs}).

The formulas for $\Sigma_{r,i}(p)$ in Eq.~(\ref{sigmaeffgeneral2}) have been
obtained  without reference to any specific gauge,
and are valid
independently of the particular choice used to compute
the quantities $\Sigma_{ab}$.  However, when the sum
in Eq.~(\ref{defaverage}) runs over all the states of the system,
the thermal averages can be written as a trace and
some simplification occurs.  In those cases
the cyclic property implies that
\begin{equation}\label{cyclicrelation}
\Sigma_{21}(p) = -e^x\Sigma_{12}(p)\,,
\end{equation}
and then,  from
Eqs.~(\ref{disprel1}) and (\ref{sigmaeffabs})
the following familiar formula follows:
\begin{eqnarray}\label{sigmaeffordinary}
\Sigma_{i}(p) & = & \frac{\mbox{Im}\,\Sigma_{11}(p)}{1 -
2n_F(x)} = \frac{\Sigma_{12}(p)}{2in_F(x)}\,,
\end{eqnarray}
where
\begin{equation}\label{nsubf}
n_F(x) = \frac{1}{e^x + 1}\,,
\end{equation}
is the fermion distribution  written in terms of the variable
%\begin{equation}\label{variablex}
$x = \beta \varepsilon - \alpha$\,,
%\end{equation}
%
with $\alpha$ being the chemical potential.

In conclusion,  the usual
expressions given in Eq.~(\ref{sigmaeffordinary})
can be applied in the Coulomb and covariant gauges,
but not in approaches like the one of LR,
where  thermal averages are not expressible as a trace
\footnote{For gauge bosons, the observation that
the relation equivalent to (\ref{cyclicrelation})
does not hold in the mixed gauge is contained
in the second paper of Ref.~\cite{landshoff}.}.
In the last  case we have to resort to
Eq.~(\ref{sigmaeffabs}),
or equivalently to Eq.~(\ref{sigmaeffgeneral2}), to determine the
absorptive part of $\Sigma_{eff}$ On the contrary, if we insist
in using  (\ref{sigmaeffordinary}), then the Landau relation as given
by  (\ref{disprel3}) or (\ref{landaurule}), is not satisfied.

It should be noticed  that,  when  the relation
of Eq.~(\ref{cyclicrelation}) is valid, then the elements $\Sigma_{ab}$
can be parametrized in terms of a single quantity $\Sigma$.
Then, Eq.~(\ref{sigmaeffgeneral}) can be written as
\begin{equation}\label{sigmaeffespecial}
\Sigma_{eff}(p) = \Sigma(p)\theta(\varepsilon) + \overline\Sigma(p)
\theta(-\varepsilon)\,,
\end{equation}
as is customarily done in the real-time formulation.
As an specific illustration of the
the general results established here,
we have considered the  the one-loop
contributions to the self-energy of  a (massless) electron
in a QED plasma, calculated
both the Coulomb and the mixed gauge \cite{villahermosa}.
In what follows we quote the main results of this calculation.

We begin with  the Coulomb gauge.
The 12 element  of the self-energy matrix is given by
\begin{equation}\label{eq3.23}
-i\Sigma_{12}(p) = (ie)(-ie)\int{\frac{d^4k}{(2\pi)^4}
i\Delta^{\mu\nu}_{21}(k)\gamma_\mu iS_{12}(p^\prime)\gamma_\nu\,,
}
\end{equation}
with $p^\prime = p + k $.  Using the  expressions for the
component $\Delta^{\mu\nu}_{21}(k)$ and $\Sigma_{12}(p)$ of the
photon an the electron bare propagators,  we find
\begin{eqnarray}\label{eq3.31}
\Sigma_{12}(p)  & = & \left(\frac{-ie^2}{4\pi^2}\right)n_F(x)
\int{\frac{d^3k}{2 \omega_k}\frac{d^3p^\prime}{2 E^{\prime}}}
(- S^{\mu\nu}\gamma_\mu\rlap / p^\prime\gamma_\nu )\times\nonumber\\
& & \left[\delta^{(4)}(p + k - p^\prime)(n_e + n_\gamma)\right.
+\delta^{(4)}(p - k - p^\prime)(1 - n_e + n_\gamma)\nonumber\\
& & + \delta^{(4)}(p - k + p^\prime)(\overline n_e + n_\gamma)
+\left.\delta^{(4)}(p + k + p^\prime)(1 - \overline n_e + n_\gamma)
\right],\nonumber\\
\end{eqnarray}
where $p^{\prime 0} = E^{\prime} = |{\vec p}\,^\prime|$ and
$k^0 = \omega_k = |\vec k|$.
$n_e$ and $n_\gamma$ stand for the electron and photon
density distributions,  while
 $\overline n_e$ is the positron distribution,
which is obtained from $n_e$ by changing
the sign of $\alpha$.  Similarly,
\begin{equation}\label{Sigma21}
-i\Sigma_{21}(p) = (ie)(-ie)\int{\frac{d^4k}{(2\pi)^4}
i\Delta^{\mu\nu}_{12}(k)\gamma_\mu iS_{21}(p^\prime)\gamma_\nu\,,
}
\end{equation}
and with the help of the relations
\begin{eqnarray}\label{s21delta12}
S_{21}(p^\prime) & = & -e^{x^\prime}S_{12}(p^\prime)\,,\nonumber\\
\Delta_{12}^{\mu\nu}(k) & = & e^{-x_\gamma}\Delta_{21}^{\mu\nu}(k)\,,
\end{eqnarray}
Eq.~(\ref{cyclicrelation}) is immediately verified at the one-loop level.
We have introduced the variables
$x^\prime  =  \beta p^\prime\cdot u - \alpha$
and
$x_\gamma  =  \beta k\cdot u = x^\prime - x$.

Turning now  the attention to
$\Sigma_{11}$, we have
\begin{equation}\label{sigma11}
-i\Sigma_{11}(p) = (-ie)^2\int\frac{d^4k}{(2\pi)^4}
i\Delta_{11}^{\mu\nu}(k)\gamma_\mu iS_{11}(p^\prime)\gamma_\nu\,.
\end{equation}
It is convenient to separate the background dependent part
$\Sigma_{11}^{(T)}$ from the standard vacuum contribution
%
%\begin{equation}\label{sigma11split}
%\Sigma_{11} = \Sigma_{11}^{(0)} + \Sigma_{11}^{(T)}\,,
%eeq
%
%where
%
\begin{equation}\label{Sigma11vac}
\Sigma_{11}^{(0)}(p) =
ie^2\int\frac{d^4k}{(2\pi)^4}\frac{(-S^{\mu\nu}\gamma_\mu
\rlap / p^\prime\gamma_\nu)}{(p^{\prime 2} + i\epsilon)(k^2 + i\epsilon)}\,.
\end{equation}
The dispersive and absorptive parts of $\Sigma_{11}^{(T)}$ are given by
\begin{eqnarray}\label{Sigma11rthermal}
\mbox{Re}\,\Sigma_{11}^{(T)}& = & e^2\int\frac{d^4k}{(2\pi)^3}
\delta(k^2)\eta_B(k)
\frac{(-S^{\mu\nu}\gamma_\mu\rlap / p^\prime\gamma_\nu)}{p^{\prime
2}}\nonumber\\
& & \mbox{} - e^2\int\frac{d^4p^\prime}{(2\pi)^3}\delta(p^{\prime 2})
\eta_F(p^\prime)
\frac{(-S^{\mu\nu}\gamma_\mu\rlap / p^\prime\gamma_\nu)}{k^2}\,,
\end{eqnarray}
\begin{eqnarray}\label{Sigma11ithermal}
\mbox{Im}\,\Sigma_{11}^{(T)} & = & \frac{e^2}{4\pi^2}
\int d^4k \delta(k^2)\delta(p^{\prime 2})
(-S^{\mu\nu}\gamma_\mu \rlap / p^\prime\gamma_\nu)
\times\nonumber\\
& & [\eta_B(k)\eta_F(p^\prime)
- \frac{1}{2}\eta_B(k) + \frac{1}{2}\eta_F(p^\prime)]\,,
\end{eqnarray}
where
\begin{eqnarray}\label{eta}
\eta_B(k) & = & n_B(x_\gamma)\theta(k\cdot u) +
n_B(-x_\gamma)\theta(-k\cdot u)\,,\nonumber\\
\eta_F(p^\prime) & = & n_F(x^\prime)\theta(p^\prime\cdot u) +
n_F(-x^\prime)\theta(-p^\prime\cdot u)\,,
\end{eqnarray}
with $n_F$ given by Eq.~(\ref{nsubf}) and
$n_B(x_\gamma)  = (e^{x_\gamma} - 1)^{-1}$.

For $\mbox{Im}\,\Sigma_{11}^{(0)}$, the Cutkosky rules yield
\begin{eqnarray}\label{Sigma11vacim}
\mbox{Im}\,\Sigma_{11}^{(0)} & = & \frac{-e^2}{8\pi^2} \int d^4k
\delta(k^2)\delta(p^{\prime 2})
 (-S^{\mu\nu}\gamma_\mu\rlap / p^\prime\gamma_\nu)\times\nonumber\\
& &[\theta(p^\prime\cdot u)\theta(-k\cdot u) +
\theta(-p^\prime\cdot u)\theta(k\cdot u)]\,,
\end{eqnarray}
and using the identity
\begin{eqnarray}\label{identity4}
2\eta_B(k)\eta_F(p^\prime) - \eta_B(k) + \eta_F(p^\prime) =
\theta(k\cdot u)\theta(-p^\prime\cdot u)
+ \theta(-k\cdot u)\theta(p^\prime\cdot u)\nonumber\\
+ (e^{x_\gamma} - e^{x^\prime})n_F(x^\prime)n_B(x_\gamma)\epsilon(k\cdot u)
\epsilon(p^\prime\cdot u),
\end{eqnarray}
it follows that, in the combination
$\mbox{Im}\,\Sigma_{11}^{(0)} + \mbox{Im}\,\Sigma_{11}^{(T)}$\,,
the vacuum term is cancelled by an identical contribution coming
from the temperature dependent  part.
The remaining terms can be rewritten by
means of  the relation
 $e^{x_\gamma}n_B(x_\gamma)n_F(x^\prime) = n_F(x)[n_F(x^\prime) +
n_B(x_\gamma)]$,
and comparing them with Eq.~(\ref{eq3.31}) it is  seen that
\begin{equation}\label{Sigma11im}
\mbox{Im}\,\Sigma_{11}(p) = \frac{i}{2}(1 - e^x)\Sigma_{12}(p)\,,
\end{equation}
in agreement with Eq.~(\ref{sigmaeffordinary}).

Introducing $\Sigma_{11\,r} \equiv \mbox{Re}\,\Sigma_{11}$,
from (\ref{Sigma11rthermal}) it follows that
\begin{eqnarray}\label{Sigma11rthermalabs}
\mbox{Im}\,\Sigma_{11r}^{(T)}(\varepsilon + i\epsilon,\vec P) & = &
\frac{-e^2}{8\pi^2}\int\frac{d^3k}{2\omega_k}\frac{d^3p^\prime}{2E^\prime}
(-S^{\mu\nu}\gamma_\mu\rlap / p^\prime\gamma_\nu)\times\nonumber\\
& & \left[\delta^{(4)}(p + k - p^\prime)(n_\gamma + n_e)\right.
+ \delta^{(4)}(p - k - p^\prime)(n_\gamma - n_e)\nonumber\\
& & + \delta^{(4)}(p -k + p^\prime)(n_\gamma + \overline n_e)
\left. +\ \delta^{(4)}(p + k + p^\prime)(n_\gamma - \overline
n_e)\right].\nonumber\\
\end{eqnarray}
In a similar  fashion, from Eq.~(\ref{Sigma11vac})
\begin{eqnarray}\label{Sigma11rvacabs}
\mbox{Im}\,\Sigma_{11r}^{(0)}(\varepsilon + i\epsilon,\vec P) & = &
\frac{-e^2}{8\pi^2}\int\frac{d^3k}{2\omega_k}\frac{d^3p^\prime}{2E^\prime}
(-S^{\mu\nu}\gamma_\mu\rlap / p^\prime\gamma_\nu)\times\nonumber\\
& & [\delta^{(4)}(p - k - p^\prime) + \delta^{(4)}(p + k + p^\prime)]\,.
\end{eqnarray}
Adding both expressions and comparing with Eq.~(\ref{eq3.31}), we arrive at
\begin{equation}\label{Sigma11rabs}
\mbox{Im}\,\Sigma_{11r}(\varepsilon + i\epsilon,\vec P)
= \frac{\Sigma_{12}(p)}{2in_F(x)}\,.
\end{equation}
This last
result explicitly shows that the physical self-energy determined by
Eq.~(\ref{sigmaeffordinary}) satisfies the Landau condition given
in Eq.~(\ref{landaurule}).  The  same conclusion remains valid in a
covariant gauge, with  $-S^{\mu\nu}$ replaced by $-C^{\mu\nu}$.

Finally, we repeat the above analysis for the approach
of LR. Our strategy  is to decompose the electron self-energy
into two pieces,
%
%\begin{equation}\label{photonpropmixed2}
%\Delta_{ab}^{\mu\nu(mix)} = \Delta_{ab}^{\mu\nu}
%+ (S^{\mu\nu} - C^{\mu\nu})\Delta_{ab}^{(0)}\,,
%\end{equation}
%
%
\begin{equation}\label{Sigmamixeddecomp}
\Sigma_{ab}^{(mix)} = \Sigma_{ab} + \Sigma'_{ab}\,,
\end{equation}
with $\Sigma_{ab}$ being the quantity we have
calculated previously
and $\Sigma'_{ab}$ representing  an additional  contribution
that arises from the term  $(S^{\mu\nu} - C^{\mu\nu})\Delta_{ab}^{(0)}$
in the photon propagator. In this way,
the 12 element of the self-energy in the mixed gauge is
expressed as in Eq.~(\ref{Sigmamixeddecomp}),
with $\Sigma_{12}$ given by Eq.~(\ref{eq3.31})  and
\begin{eqnarray}\label{sigma12}
\Sigma'_{12}(p) & = &
\frac{-ie^2}{4\pi^2}\int\frac{d^3k}{2\omega_k}\frac{d^3p^\prime}{2E^\prime}
(S^{\mu\nu} - C^{\mu\nu})\gamma_\mu\rlap / p^\prime\gamma_\nu\times\nonumber\\
 &&\left[\delta^{(4)}(p + k - p^\prime)n_e + \delta^{(4)}(p + k + p^\prime)
(1 - \overline{n}_e)\right ] .
\end{eqnarray}
By the same procedure,
\begin{eqnarray}\label{sigma21}
\Sigma'_{21}(p) &=&
\frac{-ie^2}{4\pi^2}\int\frac{d^3k}{2\omega_k}\frac{d^3p^\prime}{2E^\prime}
(S^{\mu\nu} - C^{\mu\nu})\gamma_\mu\rlap / p^\prime\gamma_\nu\times\nonumber\\
& & \left[-\delta^{(4)}(p - k - p^\prime)(1 - n_e ) - \delta^{(4)}(p - k +
p^\prime)
\overline{n}_e\right]\,.
\end{eqnarray}

For the absorptive part of $\Sigma'_{11}$, the vacuum
and the temperature dependent  contributions
can be read from Eqs.~(\ref{Sigma11ithermal}) and (\ref{Sigma11vacim})
respectively, by replacing $-S^{\mu\nu}$
by ($S^{\mu\nu} - C^{\mu\nu}$) and putting
$\eta_B(k) = 0 $.
Adding the corresponding results yields
\begin{eqnarray}\label{sigma11imixed}
\mbox{Im}\,\Sigma'_{11} & = &
\frac{-e^2}{8\pi^2}
\int{\frac{d^3k}{2\omega_k}\frac{d^3p^\prime}{2E^\prime}}
(S^{\mu\nu} - C^{\mu\nu})\gamma_\mu\rlap / p^\prime\gamma_\nu \times\nonumber\\
& & \left[\delta^{(4)}(p - k - p^\prime)(1 - n_e)
- \delta^{(4)}(p + k - p^\prime)n_e\right.\nonumber\\
& & \mbox{} - \left.\delta^{(4)}(p + k + p^\prime)(1 - \overline n_e )
+ \delta^{(4)}(p - k + p^\prime)\overline n_e\right].
\end{eqnarray}
Following similar steps for the real part of $\Sigma'_{11}$ we obtain
\begin{eqnarray}\label{sigma11rmixed}
&& \mbox{Im}\,\Sigma'_{11\,r}(\varepsilon + i\epsilon,\vec P)\  =\
\frac{-e^2}{8\pi^2}
\int{\frac{d^3k}{2\omega_k}\frac{d^3p^\prime}{2E^\prime}}
(S^{\mu\nu} - C^{\mu\nu})\gamma_\mu\rlap / p^\prime\gamma_\nu \times\nonumber\\
& & \qquad\qquad\qquad\left[\delta^{(4)}(p - k - p^\prime) (1 - n_e)
+ \delta^{(4)}(p + k - p^\prime)n_e\right.\nonumber\\
& & \qquad\qquad\qquad \mbox{} +  \left.\delta^{(4)}(p + k + p^\prime) (1 -
\overline n_e)
+ \delta^{(4)}(p - k + p^\prime)\overline n_e\right].
\end{eqnarray}
{}From the previous expressions for $\Sigma'_{ab}$, it is easy
to verify that the $\Sigma_{ab}^{(mix)}$ satisfy Eq.~(\ref{disprel1}),
and that $\Sigma_{eff}^{(mix)}$
determined from Eq.~(\ref{sigmaeffgeneral2}) or (\ref{sigmaeffabs}), satisfies
the
Landau condition Eq.~(\ref{landaurule}).
In addition,  by comparing
the formulas in Eqs.~(\ref{sigma12}) and (\ref{sigma21})  we see
that there is no simple relation between
$\Sigma'_{12}$ and $\Sigma'_{21}$, and consequently
$\Sigma_{21}^{(mix)}(p) \not = -e^x\Sigma_{12}^{(mix)}(p)$\,,
which explicitly confirms that Eq.~(\ref{cyclicrelation})
is not applicable in the mixed gauge.  In fact,
if $\Sigma_i$ were calculated by using the usual
expressions in terms of $\Sigma_{11}$
or $\Sigma_{12}$ given in (\ref{sigmaeffordinary}),
which in the present case do not yield
equivalent results,  then the resulting
formulas would not verify the Landau condition.
As already explained in the paragraph
above Eq.~(\ref{cyclicrelation}), this  is due to the fact that
the sum used to define the statistical averages
is not carried over a complete set of states of the system.

\end{document}